\documentclass[journal]{IEEEtran}

\newcommand{\ignore}[1]{}

\usepackage{cite}
\usepackage{algorithmic}
\usepackage[ruled, vlined]{algorithm2e}
\usepackage{xcolor}
\usepackage{multirow}
\usepackage{textcomp}
\usepackage{booktabs}

\ifCLASSINFOpdf
   \usepackage[pdftex]{graphicx}
   \graphicspath{{./pdf/}{./jpeg/}}
   \DeclareGraphicsExtensions{.pdf,.jpg,.png}
\else
\fi
\usepackage{amsmath}
\usepackage{amssymb, amsthm}

\theoremstyle{definition}

\begin{document}
\title{A Context-Aware Readout System \\for Sparse Touch Sensing Array \\Using Ultra-low-power Always-on Event Detection}

\author{Hyeri~Roh,~\IEEEmembership{Graduate Student Member,~IEEE},
        and~Woo-Seok~Choi,~\IEEEmembership{Member,~IEEE}% <-this % stops a space
\thanks{This work was supported in part by the New Faculty Startup Fund from Seoul National University and by National R\&D Program through the National Research Foundation of Korea (NRF) funded by Ministry of Science and ICT (2020M3H2A1078119). 
}% <-this % stops a space
\thanks{H. Roh and W.-S. Choi are with the Department
of Electrical and Computer Engineering and the Inter-University Semiconductor Research Center, Seoul National University, Seoul 08826, South Korea (e-mail: hrroh@snu.ac.kr, wooseokchoi@snu.ac.kr).}% <-this % stops a space
}


\markboth{Journal of \LaTeX\ Class Files,~Vol.~14, No.~8, August~2015}%
{Shell \MakeLowercase{\textit{et al.}}: Bare Demo of IEEEtran.cls for IEEE Journals}



\maketitle
% \fi
\IEEEpeerreviewmaketitle

%%%%%%%%%%%%%%%%%%%%%%%%%%%%%%%%%%%%%%%%%%%%%%%%%%%%%%%%%%%%%%%%%%%%%
% Abstract

\begin{abstract}
Increasing demand for larger touch screen panels (TSPs) places more energy burden to mobile systems with conventional sensing methods.
To mitigate this problem, 
taking advantage of the touch event sparsity,
this paper proposes a novel TSP readout system that can obtain huge energy saving by turning off the readout circuits when none of the sensors are activated.  
To this end, a novel ultra-low-power always-on event and region of interest detection based on lightweight compressed sensing is proposed.
Exploiting the proposed event detector, the context-aware TSP readout system, which can improve the energy efficiency by up to {42$\times$}, is presented.
\end{abstract}


%%%%%%%%%%%%%%%%%%%%%%%%%%%%%%%%%%%%%%%%%%%%%%%%%%%%%%%%%%%%%%%%%%%%%
% Introduction

\section{Introduction}
\label{sec:intro}

\IEEEPARstart{C}{apacitive} sensors have been used for decades, recognized for its stability of operation in temperature variations and high sensing accuracy in various fields including touch screen panel (TSP) technology.
Many products such as mobile phones, tablets, and TVs take advantage of TSPs,
and the demand for larger and faster TSPs keeps increasing.
However, larger TSPs increase the number of sensors in an array and requires faster readout circuit,
both of which exacerbate the energy dissipation.
Since energy efficiency is a huge concern for mobile systems, this calls for a new technique to achieve the excellent energy efficiency with large TSPs.
Noticing that turning on the readout circuits all the time is energy consuming and that only a small fraction of sensors is activated for most TSP applications,
this paper proposes a novel TSP readout system that can obtain huge energy saving by turning off the readout circuits when none of the sensors are activated.
To this end, this paper proposes a novel ultra-low-power always-on event, or region-of-interest, detection using lightweight compressed sensing.
Moreover, exploiting the proposed event detector implemented in RTL and synthesized in a 28\,nm process, this paper presents a context-aware TSP readout system, which can improve the energy efficiency by up to {42$\times$} over conventional systems.

%%%%%%%%%%%%%%%%%%%%%%%%%%%%%%%%%%%%%%%%%%%%%%%%%%%%%%%%%%%%%%%%%%%%%
% Background

\section{Background}
\label{sec:background}

\subsection{Touch Sensor Array \& Readout Circuits}

\begin{figure}[t]
\centering
\includegraphics[width=0.9\columnwidth]{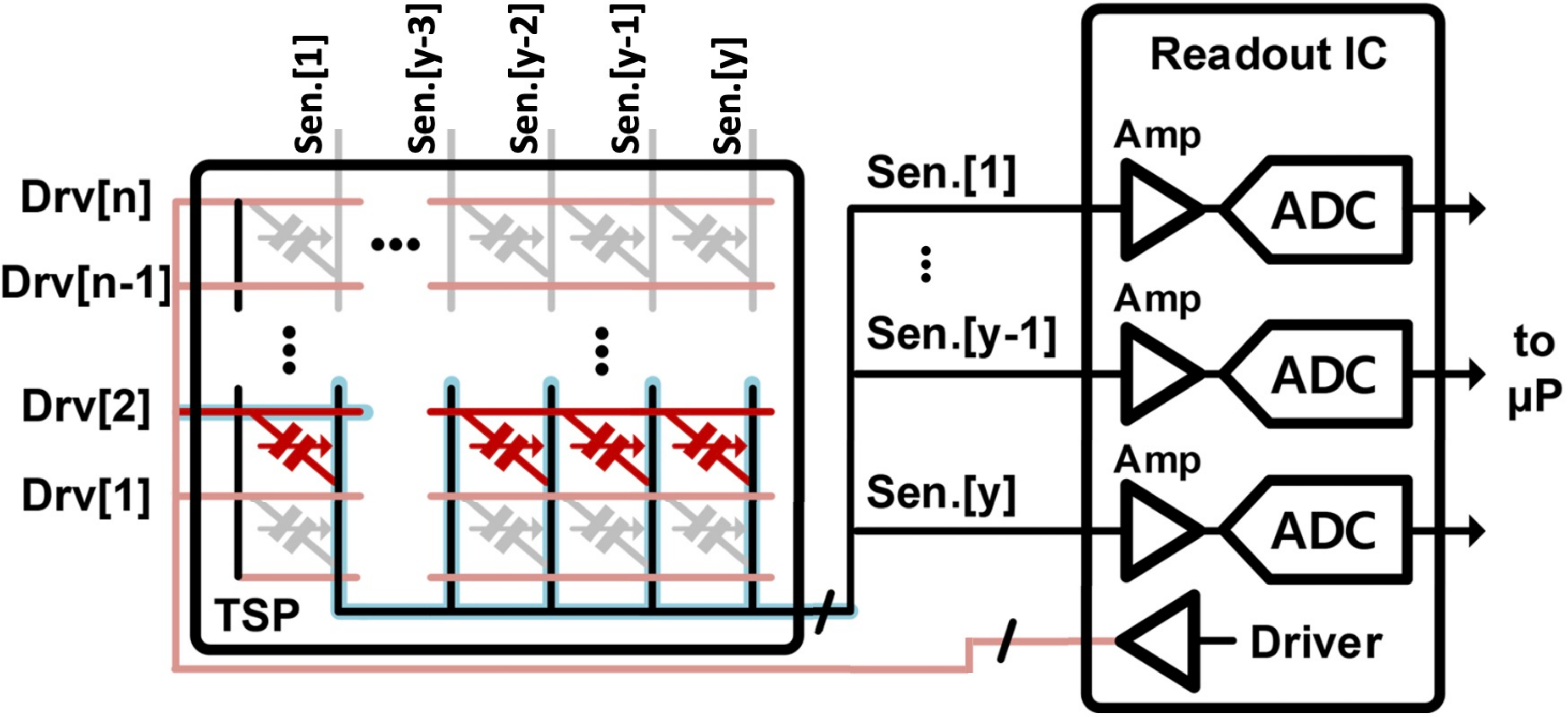}
\vspace{-0.8em}
\caption{{Overall block diagram of TSP and readout circuits.}}
\label{fig:TSP}
\vspace{-1em}
\end{figure}

Fig.~\ref{fig:TSP} shows the overall block diagram of a TSP and interface readout circuits. 
The TSP is composed of multiple driving and sensing channels. 
The touch information is embedded in the mutual capacitance $C_M$, which is the parasitic capacitance between the driving channel and the sensing channel.
$C_M$ under the region, where fingers are touched, is reduced, and 
the driver applies an excitation signal to each driving channel, Drv[i], to measure $C_M$. 
The excitation signals modulated by $C_M$ are transferred through the sensing channels to the readout circuit, where the small input signal is amplified and digitized.
The main components in the readout integrated circuits (ICs) are the charge amplifier, the analog-to-digital converter (ADC), and the driver, and the amount of power consumed by these components depends on the performance metric such as frame rate and signal-to-noise ratio (SNR). 
Although each block can be designed to minimize the power individually while satisfying the required performance, 
there is fundamental trade-off between the frame rate, SNR, and power consumption~\cite{razavi2002design}. 
Better performance, e.g. higher frame rate and SNR, inevitably incurs higher power consumption. 

Although the circuit-level design techniques have improved the performance significantly, these techniques alone are not sufficient to support the increasing demand for TSPs with larger sizes because the number of sensors and the power of the readout circuits increase accordingly. 
This is especially undesirable for systems like mobile devices with severe energy constraints. 
This problem can be mitigated with the help of various system-level approaches, among which how to access and read multiple sensors and how to extract the wanted information will be introduced in the next section. 

%%%%%%%%%%%%%%%%%%%%%%%%%%%%%%%%%%%%%%%%%%%%%%%%%%%%%%%%%%%%%%%%%%%%%
% Prior art

\subsection{Existing Sensing Techniques}
\label{sec:prior_art}

\begin{figure}[t]
\centering
\includegraphics[width=\columnwidth]{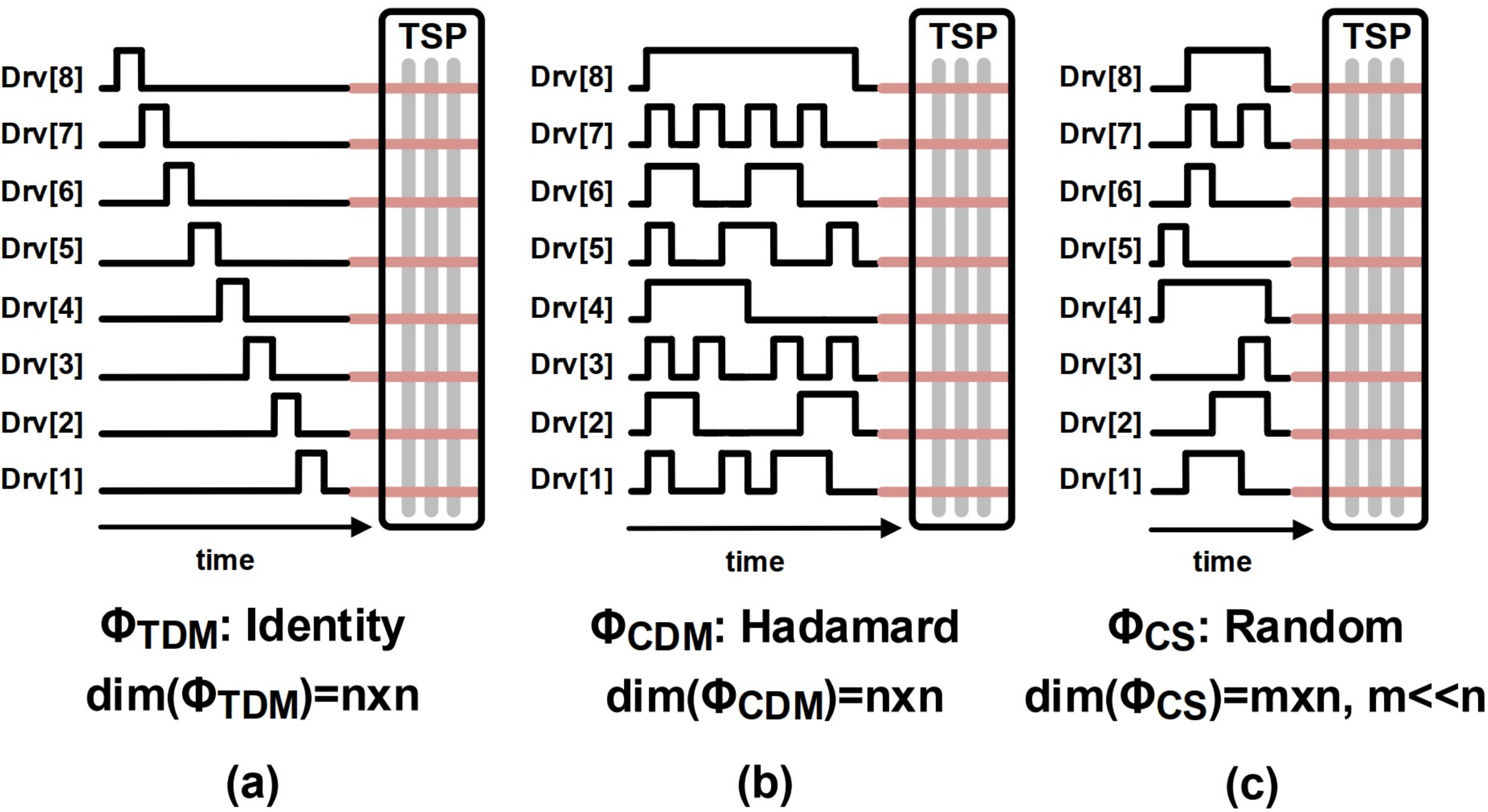}
\vspace{-1.9em}
\caption{Existing TSP sensing schemes: (a) TDM, (b) CDM, and (c) CS.}
\label{fig:prior_art}
\vspace{-1em}
\end{figure}

{Since each column-wise signal path reads $C_M$'s modulated by the driving signals, 
the measured values by ADCs can be expressed as: $\mathbf{y_i}  = \mathbf{\Phi x_i + n},$  where $\mathbf{x_i}$ is mutual capacitances in the column $\mathbf{i}$ ($\mathbf{x_i} \in \mathbb{R}^n$, $\mathbf{i} \in \{1, ..., y\}$ in Fig.~\ref{fig:TSP}).
$\mathbf{\Phi}$ is the sensing (measurement) matrix determining how to modulate the driving channel, and $\mathbf{n}$ represents circuit random noise.}

\subsubsection{Time Division Multiplexed Sensing (TDM)}

The simplest way to read out data from a sensor array is to use the TDM scheme~\cite{park2014reconfigurable}  
(see Fig.~\ref{fig:prior_art} (a)). 
$\mathbf{\Phi_{TDM}}$ is the $n \times n$ identity matrix, so it requires no extra signal recovery phase, i.e. $\mathbf{\hat{x_i}} = \mathbf{y_i}$ for each column.
Despite its simplicity,
the energy efficiency of TDM becomes severely degraded as the number of sensors increases. 
For instance, if the number of sensors becomes twice while the frame rate is fixed, the allocated time for reading each sensor becomes half.
Then, the power consumption of the readout circuits becomes more than twice for 2$\times$ higher bandwidth while maintaining the same SNR. 
Thus, when the TSP size or when the required frame rate is large (around hundreds Hz), the overall energy consumption of TDM becomes unacceptable in mobile systems.

\subsubsection{Code Division Multiplexed Sensing (CDM)}

Applying CDM for reading out the touch sensor array data has been introduced in \cite{park2016pen,park2019power}. 
{$\mathbf{\Phi_{CDM}}$ is an $n \times n$ orthogonal matrix, and since all the orthogonal matrices are invertible and the inverse is its transpose, the sensor data can be recovered by $\mathbf{\hat{x_i}} = \mathbf{\Phi_{CDM}}^\intercal \mathbf{y_i}$.}
In theory, any orthogonal matrix can be used, but if the elements of $\mathbf{\Phi_{CDM}}$ are composed of +1/0/-1, then it becomes simple to implement in hardware.
For instance, Hadamard matrix is used in \cite{park2016pen}. 
CDM helps reduce noise in the signal recovery phase and achieve higher SNR if the noise is independent and identically distributed (i.i.d.).

However, like TDM, CDM also should measure as many samples as the number of sensors, which increases the power of the amplifiers and ADCs with large TSPs.
Moreover, the dynamic power of the driver in Fig.~\ref{fig:TSP} also increases dramatically 
because the CDM driver should keep switching during the whole sensing period (see Fig.~\ref{fig:prior_art} (b)) and the parasitic capacitance of the driving channels increases as the TSP size gets larger. 
Therefore, CDM also suffers from degraded energy efficiency with large TSPs.

\subsubsection{Compressed Sensing (CS)}

CS has been introduced to read out sparse data from sensor arrays~\cite{DonohoTIT2006}.
The distinct feature of CS is that it does not require as many samples as TDM or CDM does, 
thereby making it possible to save huge amount of energy consumed by the readout circuits.
$\mathbf{\Phi_{CS}}$ is an $m \times n$ rectangular matrix with $m < n$, yielding an under-determined recovery equation.
Still, if $\mathbf{x_i}$ is sparse, 
it is possible to recover $\mathbf{x_i}$ using properly chosen $\mathbf{\Phi_{CS}}$.
In practice, however, there are two main difficulties that prevent CS from being used for mobile systems.

First, to take advantage of CS, we need to construct a measurement matrix $\mathbf{\Phi_{CS}}$ that satisfies the Restricted Isometry Property (RIP)~\cite{candes2005decoding}. 
However, constructing such matrices is hard, 
so most of the measurement matrix $\mathbf{\Phi_{CS}}$ is built using random elements.
Dense random matrices, where the elements are generated by an i.i.d. Gaussian or Bernoulli process, are widely used in CS because they satisfy RIP with high probability~\cite{baraniuk2008simple}. 
Due to simplicity, Bernoulli matrices composed of binary elements, +1/-1, are preferred, 
but still they require an on-chip random seed or a large memory to store them.
In order to apply CS for TSPs with small memory overhead, 
Bernoulli matrix and its circulant version were used in \cite{luo2012compressive,luo2014low}.
Pursuing lower complexity for implementation, 
\cite{amini2011deterministic} exploited deterministic matrices for CS.

The bigger issue of using CS in mobile systems stems from the complex signal recovery algorithms.
Using $\ell_1$ minimization, which provides theoretical guarantees for signal recovery~\cite{needell2008greedy}, 
requires huge computation and time. 
Due to computational simplicity, greedy algorithms are attractive,
and the variants were proposed for TSPs in \cite{luo2012compressive,luo2014low}.
However, they still suffer from either high power or low detection probability.
Even if we can save huge energy of the readout circuits thanks to CS, 
the overall system does not gain any benefits due to the computational energy for signal recovery. 
This has limited the usage of CS to the applications, where signal recovery can be done offline.
Thus, conventional CS is not adequate for large TSPs, where not only signal sensing but its recovery should be done in an energy-efficient manner. 

%%%%%%%%%%%%%%%%%%%%%%%%%%%%%%%%%%%%%%%%%%%%%%%%%%%%%%%%%%%%%%%%%%%%%
% Proposed

\vspace{-0em}
\section{Proposed TSP Readout System}
\label{sec:proposed}

\subsection{Overall System Description}

\begin{figure}[t]
\centering
\includegraphics[width=\columnwidth]{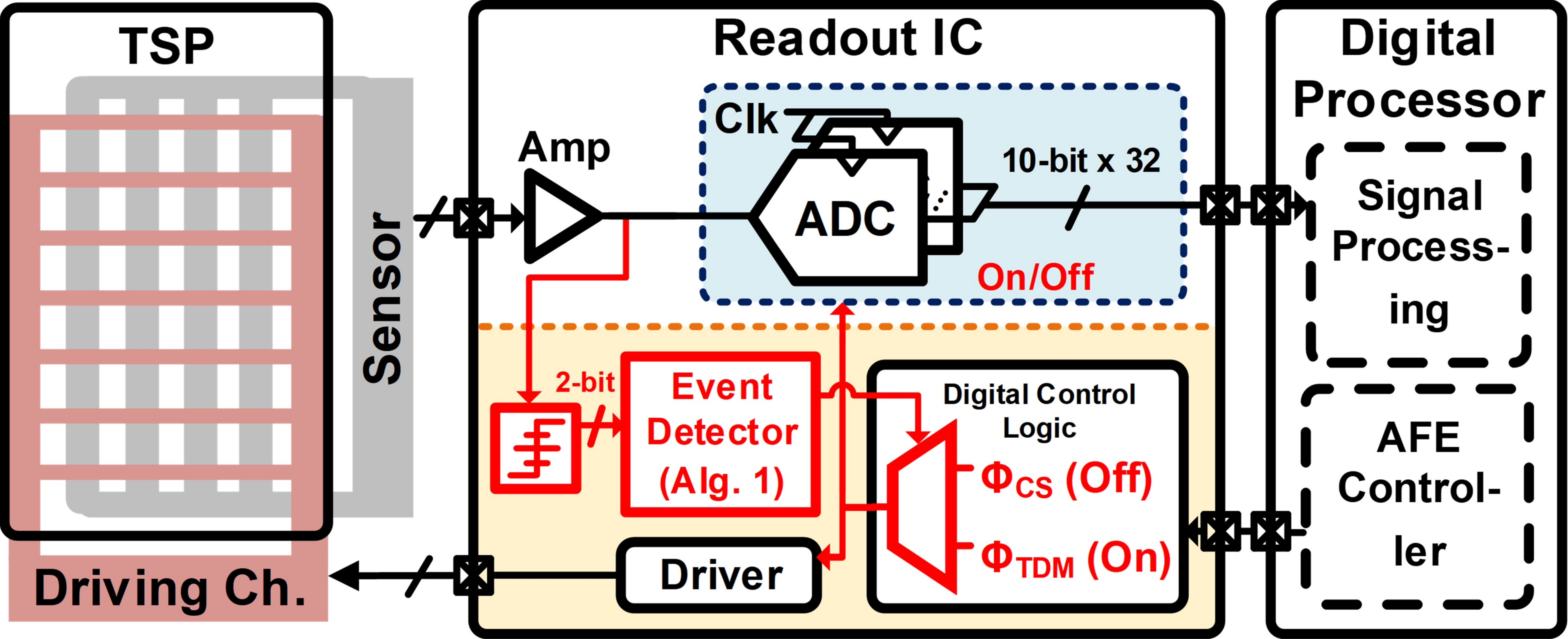}
\vspace{-1.7em}
\caption{{Overall block diagram of the proposed system. The blocks added for context-aware operation are marked in red.}}
\label{fig:proposed}
\vspace{-1.4em}
\end{figure}

High-resolution ADCs are typically the most energy-consuming block in conventional TSP readout systems, and since the touch events are sparse in time, great energy saving can be gained if the ADCs are turned off most of the time, and operate only when needed. 
To build such systems, we propose an ultra-low-power sparse event detection scheme with negligible hardware overhead. 
With the proposed system, energy saving can be achieved as long as the touch events happen sparsely and the high-resolution ADC is the dominant energy-consuming block in the system.

Fig.~\ref{fig:proposed} shows the overall block diagram of the proposed TSP readout system
with the following distinct features: 1) ultra-low-power always-on region-of-interest (ROI), or event detection based on CS, and 2) context-aware readout operation by sensing only the regions where the events are detected.

For the event detection, multiple sensors are read simultaneously using the proposed measurement matrix.
To discard the need for random number generators and reduce the required memory,
the system exploits a deterministic measurement matrix.
Furthermore, all the elements in the matrix are composed of either +1, 0, or -1, for hardware simplicity.
Taking advantage of the sparsity nature of the touch event, 
the proposed matrix enables detecting the event and ROI with a simple algorithm, which can be implemented with ultra-low power.

With the help of the ultra-low-power event detection,
the proposed sensing system can achieve \textit{energy-proportional operation}, or \textit{context-aware operation}~\cite{barroso2007case,VenkataramaniDATE2015},
which allows the sensing system to figure out the context of sensor data and selectively transmit the data, e.g. activated sensor data only, to the processor.
In contrast to conventional systems, the overall power consumption of the proposed one is no longer directly proportional to the total number of sensors but rather proportional to the occurrence of events of interest.
For such context-aware systems, since communication between the readout IC and the processor is required only when events occur, communication link I/Os need to support rapid turning on/off operation for overall system energy proportionality~\cite{choi2015burst,shu201623,kim202012}.
\begin{figure*}[t]
\centering
\includegraphics[scale=0.2]{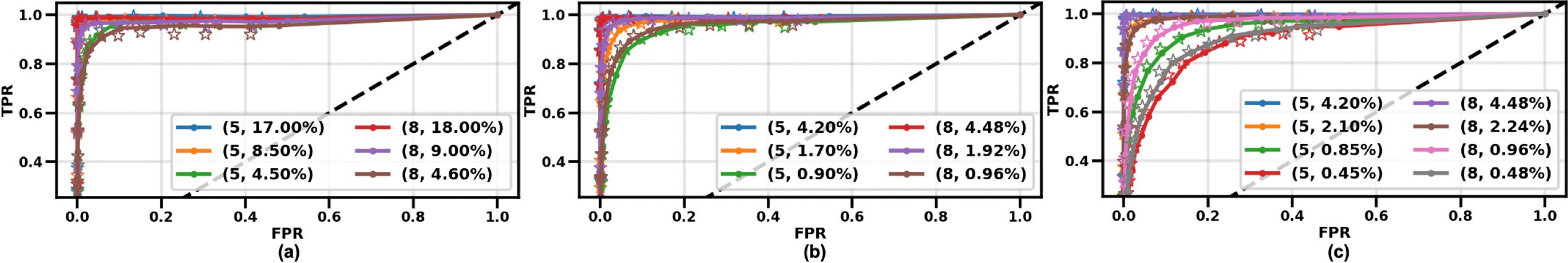}
\vspace{-0.5em}
\caption{{ROC curves with various sets of parameters ($k$, sampling ratio) : (a) $N=1000$, (b) $N=5000$, (c) $N=10000$.}}
\label{fig:10000_5000}
\vspace{-1em}
\end{figure*}
The proposed context-aware readout operation is performed in two steps: 1) event or ROI detection using CS and 2) ROI readout using TDM (see {the yellow area of} Fig.~\ref{fig:proposed}).
The system activates only the ultra-low-power event detector to detect the region of activation (described in Section~\ref{sec:proposed_detection})
and then only when detected the sensors in that region are read sequentially using TDM {by the high-resolution ADCs (the blue area of Fig.~\ref{fig:proposed}).}
This makes the system power consumption proportional to the number of activated events, not to the number of sensors in a TSP.
Since in practice the touch events are sparse not only in location but also in time,
the proposed system can save huge amount of power, consuming nearly zero power during when no event happens.
Moreover, since only the activated sensors are read sequentially using TDM, 
the complex signal recovery algorithm by which conventional CS suffers from huge energy consumption can be avoided.
Note that the ultra-low-power event detector plays a key role in the proposed system, which will be described next.

\vspace{-0.7em}
\subsection{Proposed Event Detection}
\label{sec:proposed_detection}
 
The proposed deterministic measurement matrix $\mathbf{\Phi}_k$ for sparse touch event detection is constructed as follows.
Let $\mathbf{e}_i \in \mathbb{Z}^m$ denote a column vector whose $i$-th element is 1 and all the other elements are zero, 
and let $\mathbf{a}_{i,k} \in \mathbb{Z}^m$ denote a column vector whose elements in $i$-th to $(i+k)$-th positions are -1 and all the others are zero. In other words, 
$\mathbf{a}_{i,k} = - \sum_{j=i}^{i+k} \mathbf{e}_j$.
Then the columns of $\mathbf{\Phi}_k$ are composed of $\mathbf{e}_i$'s and $\mathbf{a}_{i,k}$'s as described below. 
(For simplicity, suppose that $m$ is an integer multiple of $k$ in which case $n = m + \frac{m}{k}$.) 
\vspace{-0.5em}
\renewcommand{\arraystretch}{0.7}
\begin{equation*}
	\mathbf{\Phi}_k  = \left[\arraycolsep=2.4pt
	\begin{array}{cccc|cccc}
						\vrule & \vrule &  & \vrule & \vrule & \vrule & & \vrule \\
						\mathbf{e}_1 & \mathbf{e}_2 & \dots & \mathbf{e}_m & \mathbf{a}_{1,k} & \mathbf{a}_{k+1,k} & \dots & \mathbf{a}_{m-k+1,k}  \\
						\vrule & \vrule &  & \vrule & \vrule & \vrule & & \vrule
						\end{array} 	
					\right]
\end{equation*}
\renewcommand{\arraystretch}{0.8}
For example, when $k$ is 2, $\mathbf{\Phi}_2$ can be expressed as
\[
	\mathbf{\Phi}_2	 = \begin{bmatrix}
					1 & 0 & 0 &\dots & 0 & -1 & 0 & \dots & 0\\
					0 & 1 & 0 & \dots & 0 & -1& 0 & \dots & 0\\
					0 & 0 & 1 & \dots & 0 & 0 & -1 & \dots & 0\\
					\vdots & \vdots & \vdots & \ddots & \vdots & \vdots  & \vdots & \ddots & \vdots \\
					0 & 0 & 0 & \dots & 1 & 0 & 0 & \dots & -1
					\end{bmatrix}.	
\]

The sensing matrix $\mathbf{\Phi}_k$ is designed using ternary elements $\{-1, 0, 1\}$ such that any $k$ or less than $k$ columns of the matrix are linearly independent.
This guarantees that any $k$-sparse signals can be detected by monitoring whether the measurements are zero (noiseless) or close to zero (in the presence of noise).
Since the signal of interest is always positive in TSP, the index of the activated sensors can be identified by checking the sign of the measurements and their magnitude.
Note that since $m < n$, less number of measurements is required to recover $k$-sparse signals from $n$ sensors. 
In other words, with $\mathbf{\Phi}_k$, 
the sampling ratio is $\frac{k}{k+1}$ ($m$ samples from $n=m+\frac{m}{k}$ sensors). 
Compared to conventional CS, the sampling ratio with $\mathbf{\Phi}_k$ has the following downsides:
1) it does not approach zero asymptotically as $n$ increases, and 2) it approaches 1 as $k$ increases, implying that the required number of samples is almost the same as the number of sensors like TDM or CDM.
\setlength{\parskip}{0\baselineskip}
However, the structured nature of $\mathbf{\Phi}_k$ provides a very simple way to detect the active sensors. 
We first search negative measurements and find the active sensors in the index $[m+1, n]$.
If none of the sensors in the index 
$[1, m]$ are active, any non-zero measurements should appear in $k$ consecutive indexes. 
If negative values do not appear consecutively, we can easily find out which sensors in the index $[1, m]$ are active.
Note that only simple operations such as comparison or addition are required for detection, allowing low energy consumption.
One may think that detecting ROI would be possible by driving multiple sensors simultaneously and comparing the output with a certain threshold.
However, the sensor output is composed of a large DC and a small varying component that needs to be detected, and simply driving multiple sensors with one adds up the DC components of the sensor outputs, which greatly reduces the dynamic range of the analog circuits such as amplifiers and ADCs. 
To prevent this, the sensing matrix $\mathbf{\Phi}_k$ is designed such that the elements in each row are summed to zero and the DC components of the sensor outputs can be canceled out each other.

\setlength{\textfloatsep}{0em}
\begin{algorithm}[t]
\DontPrintSemicolon
\SetArgSty{textnormal}
\KwIn{2-bit quantized measurements $\mathbf{y} = Q(\mathbf{\Phi}_{k,l}\mathbf{x+n)} \in \mathbb {R}^m$}
\KwOut{Indicator vector $\mathbf{w} \in \mathbb{R}^N$ \\ ($w_i = 1$ if $x_i$ is activated, otherwise 0)}
  \textbf{Parameter} Vth: threshold level for input quantization
  Initialize $\mathbf{w} = 0 $; \quad $i = 1$\;
 \For{$i \leq m$}{
  \If{$\exists y_j <  \mathrm{-Vth \quad for} \ j \in [i, i+k-1]$}{
    $w_{l(m+\lceil i/k \rceil)-l+1} = \dotsi = w_{l(m+\lceil i/k \rceil)} = 1$\;
    \If{$y_j > -\frac{\mathrm{Vth}}{2} \quad \mathrm{for} \ j \in [i, i+k-1]$}  {
    $w_{lj-l+1} = \dotsi = w_{lj} = 1$
    }
 }
 \Else{
   \If{$y_j > \mathrm{+Vth \quad for} \ j \in [i, i+k-1]$}  {
    $w_{lj-l+1} = \dotsi = w_{lj} = 1$
    }
 }
 $i = i + k$
 } 
 \caption{Detecting ROI with $\mathbf{\Phi}_{k,l}$ from noisy measurements.}
 \label{alg2}
\end{algorithm}
 
Using the structure of $\mathbf{\Phi}_k$, a new measurement matrix $\mathbf{\Phi}_{k,l} \in \mathbb{Z}^{m \times nl}$ with the improved sampling ratio is constructed by repeating each column of $\mathbf{\Phi}_k$ $l$ times, 
where $l$ is a design parameter chosen based on the sparsity $k$ and the total number of sensors $N$.
$k$ and $N$ are decided by the system specifications, while $l$ and $m$ should be chosen by a designer to satisfy $N \leq nl$, where $n=m+\frac{m}{k}$.
If $N < nl$, some columns of $\mathbf{\Phi}_{k,l}$ are repeated $(l-1)$ times, rather than $l$ times
to make the $\mathbf{\Phi}_{k,l}$ size $m \times N$.
This construction allows us to think of $l$ (or $l-1$) chunks of sensors as a single sensor, and 
we can apply the same detection algorithm to find out which sensor chunks are activated.
In other words, the $k$-sparse signal can be detected from $N=n \times l=l(m+\frac{m}{k})$ sensors using only $m$ samples, i.e. sampling ratio being $\frac{k}{l(k+1)}$.
The event detection process with $\mathbf{\Phi}_{k,l}$ works the same as the one with $\mathbf{\Phi}_{k}$, except that $l$ sensors in a chunk are treated as one sensor.
When a sensor is detected active with $\mathbf{\Phi}_{k,l}$, all $l$ sensors in the chunk are considered candidates for being active. 

\setlength{\parskip}{0\baselineskip}
Taking noise into account, the measured values $\mathbf{\Phi}_{k,l}\mathbf{x+n}$ are compared with properly chosen thresholds (Vth) and quantized into 4 levels:
$(-\infty, -\textrm{Vth}),$ $(-\textrm{Vth}, -\frac{\textrm{Vth}}{2}),$ $(-\frac{\textrm{Vth}}{2}, \textrm{Vth})$, and $(\textrm{Vth}, \infty)$.
The quantized measurements are processed using Alg.~\ref{alg2} for ROI detection.
We first search for the location of the large negative samples in the measurement (corresponding to first if-statement in Alg.~\ref{alg2}). This provides the information about the index of the activated sensors.
Then, the $k$ consecutive samples associated with the sensors considered activated are checked whether some of the samples are close to zero.
If so, this provides additional information about the index of other activated sensors (corresponding to second if-statement in Alg.~\ref{alg2}).
If none of the measured samples are negative with large magnitude, none of the last $\frac{m}{k}$ sensors are activated.
Then, we search for the positive samples with large magnitude and identify the index of the activated sensors (corresponding to else-statement in Alg.~\ref{alg2}).
\ignore{Note that, considering $l$ chunks as a single sensor, Alg.~\ref{alg2} is basically the same as the simple detection algorithm with $\mathbf{\Phi}_k$, Alg.~\ref{alg2}.}
Only the sensors in the detected chunks are read sequentially using TDM.
\setlength{\parskip}{0\baselineskip}

%%%%%%%%%%%%%%%%%%%%%%%%%%%%%%%%%%%%%%%%%%%%%%%%%%%%%%%%%%%%%%%%%%%%%%
% Experiments
\setlength{\parskip}{0\baselineskip}
\vspace{-0.5em}
\section{Experiments}
\label{sec:experiments}

Ideally, the proposed system should only read the sensors in the activated chunks. 
However, due to noise, inactive chunks could be falsely detected as active. 
Note that false alarm incurs energy overhead since inactive sensors are read unnecessarily and it should be minimized.

We evaluated Alg.~\ref{alg2} with different values of sparsity $k$ (5, 8) and total number of sensors $N$ (1000, 5000, 10000)\footnote{
{
A high-resolution TSP such as the one reported in \cite{miura201412} with a 1\,mm-pitch allows more natural human interfaces and highly accurate expressions, which are required for a variety of applications.
Scaling up the 248 sensors with a 6.25\,mm pitch presented in \cite{miyamoto2014143} to a 1\,mm pitch leads to 1550 sensors in each column.
This would result in using up to $10^3$-$10^4$ sensors in the future as the TSP size increases proportionately.}
}.
\begin{figure}[t]
\centering
\includegraphics[width=\columnwidth]{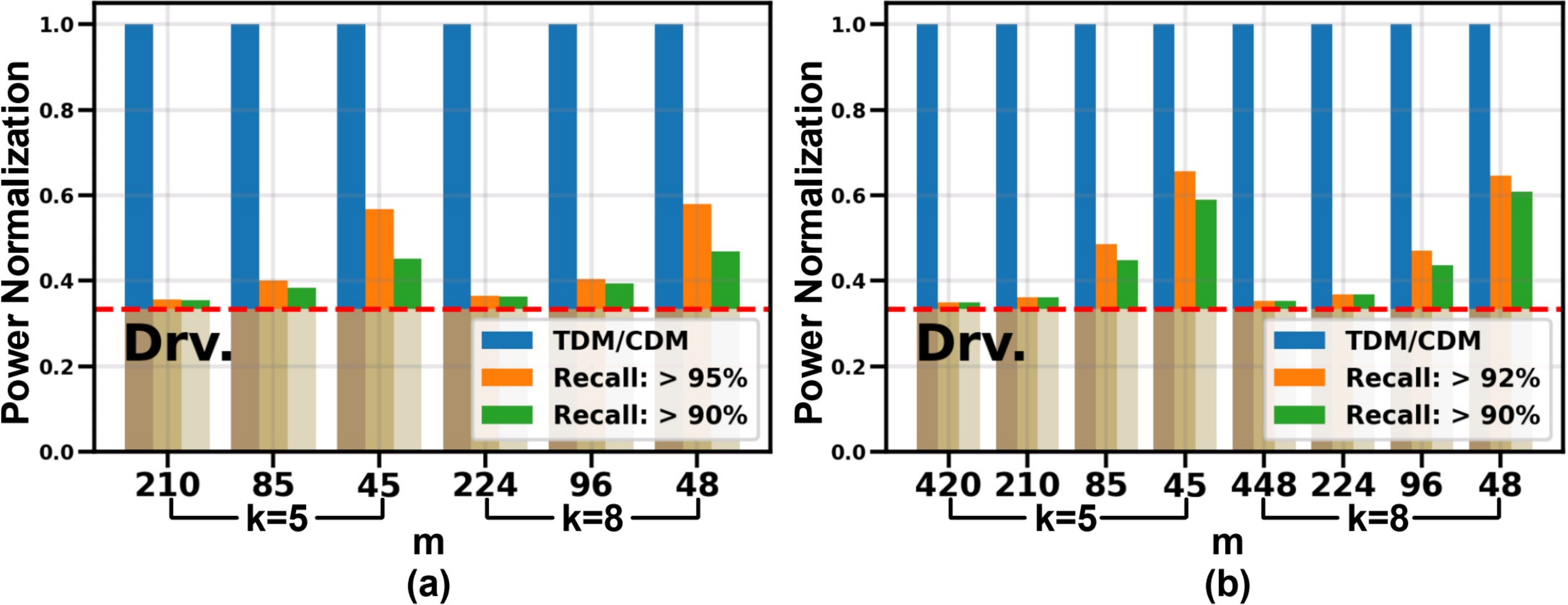}
\vspace{-2.3em}
\caption{Simulated energy saving with various sets of parameters: (a) $N=5000$, (b) $N=10000$.}
\label{fig:10000_5000_power}
\vspace{0.5em}
\end{figure}
For each set of $k$ and $N$, different number of samples $m$ and number of sensors in a chunk $l$ were chosen.
Then, for each parameter set, Vth in Alg.~\ref{alg2} was swept to obtain various receiver operating characteristic (ROC) curves, plotted in Fig.~\ref{fig:10000_5000}.
Fig.~\ref{fig:10000_5000} (a) was obtained with 25\,dB TSP SNR and 40\,dB readout circuits SNR while (b) and (c) were with 30\,dB SNR for TSP~\cite{patrick2011}.
When $N$=1000 with 30\,dB TSP SNR, all cases of $k$ and sampling ratio showed the perfect ROC curves,
so lower TSP SNR of 25\,dB was tested when $N$=1000.
The star markers in Fig.~\ref{fig:10000_5000} show the ROC of a slightly modified detection algorithm, where additional post-processing is performed to reduce the false positive rate (FPR), e.g. considering the chunk is falsely activated if the chunk contains both above-(+Vth) and below-(-Vth) values.
From Fig.~\ref{fig:10000_5000}, we can see that the sampling ratio around 4.2\,\% in 30\,dB and 40\,dB SNR shows the perfect detection, while true positive rate (TPR) and FPR get worse as the sampling ratio reduces.

The proposed sparse event detector that runs Alg.~\ref{alg2} was synthesized in a 28\,nm CMOS process. 
It is implemented with compact combinational logic blocks composed of counters, comparators, and logic gates taking 2-bit (coarsely quantized) inputs and generating as output a binary vector. It indicates whether the event is detected or not at the corresponding cell.

The logic area depends on the maximum frame-rate and the sparsity parameter $k$, which are decided by the system or application requirement. 
The operating clock frequency is set to meet the maximum TSP frame rate 200\,Hz.
The detector is always on with minimal power overhead, and all the readout circuits can be turned off unless any event is detected.
Based on \cite{park2019power}, power models for an amplifier, ADC, and driver were built:
0.3\,mW per channel for driver, 0.2\,mW per channel for amplifier, and 0.4\,mW per channel for ADC at\ignore{250\,kHz} 120\,Hz frame rate.
We adopted it for the existing circuits, and the area and power overhead of the proposed digital block were obtained from the RTL synthesis result.
Fig.~\ref{fig:10000_5000_power} shows 2.14\,-\,42.47\,$\times$ energy saving of the proposed readout system over conventional sensing methods with recall $>$ 90\,\%.

\begin{table}
\setlength{\tabcolsep}{4pt}
\centering
\caption{{Area Overhead and Power Saving Gain of the Proposed Context-Aware Readout Operation.}}
\vspace{-0.5em}
\label{areaPower}
{\small
\begin{tabular}{@{}c@{}c@{}c@{}ccc@{}}
\toprule[1.5pt]
& \multicolumn{2}{c}{\textbf{Prop. Detector}} & \textbf{~\cite{park2019power}} & \textbf{~\cite{park20160}} & \textbf{~\cite{an2021readout}} \\
\midrule[0.2pt]
\midrule[0.2pt]
Process & \multicolumn{2}{c}{28\,nm} & 180\,nm & 350\,nm & 130\,nm \\ [0.5pt]
\textbf{Scaled area} &  \multicolumn{2}{c}{\textbf{66.61$\boldsymbol{\mathrm{\mu m^2}}$}} & \textbf{0.87$\boldsymbol{\mathrm{mm^2}}$} & \textbf{0.04$\boldsymbol{\mathrm{mm^2}}$} & \textbf{3.35$\boldsymbol{\mathrm{mm^2}}$} \\ [0.7pt]
{Power} & \multicolumn{2}{c}{w/o detector} & {0.6} & {0.25} & {1.49} \\
(mW/channel)& \multicolumn{2}{c}{w/ detector} & {0.014-0.28} & {0.007-0.12} & {0.047-0.72} \\[0.8pt]
{\textbf{Power saving}} & \multicolumn{2}{c}{-} & \textbf{2.1-42.5$\boldsymbol{\mathrm{\times}}$} & \textbf{2.1-33.7$\boldsymbol{\mathrm{\times}}$} & \textbf{2.1-31.7$\boldsymbol{\mathrm{\times}}$} \\
\bottomrule[1.2pt]
\end{tabular}
}
\end{table}

Similarly, we also quantified how much power saving can be obtained from other works using the proposed detector, and the results demonstrate a significant reduction in power consumption.
It saves power of 2.06\,-\,33.66\,$\times$ in \cite{park20160} and 2.06\,-\,31.71\,$\times$ in \cite{an2021readout} using the proposed context-aware readout operation. 
The proposed detector occupies 33.23\,$\mu$m$^2$, 52.42\,$\mu$m$^2$, and 66.61\,$\mu$m$^2$ for $k$=5, 8, and 10, respectively.
In comparison, the active areas of the readout circuits (including analog front-end and high-resolution ADCs) in \cite{park2019power}, \cite{park20160}, and \cite{an2021readout} are 36\,mm$^2$, 5.52\,mm$^2$, and 72.25\,mm$^2$, respectively.
Even with perfect scaling to a 28 nm process, they are scaled down to 0.87\,mm$^2$, 0.04\,mm$^2$, and 3.35\,mm$^2$.
On the other hand, the proposed detector occupies only 66.61\,$\mu$m$^2$ when $k$=10, which indicates that the area overhead of the proposed detector is negligible.
Table~\ref{areaPower} summarizes the area overhead and power-saving gain of the proposed system.

Since the same number of measurements is read for TDM and CDM, the amplifier bandwidth and ADC sampling rate should be the same when the frame rate and the number of sensors are both the same. 
This results in the same power usage for TDM and CDM.
On the other hand, the proposed context-aware readout system power is divided into two parts, 
the sparse event detector and the sequential readout, 
using the power models and the simulated TPR/FPR, the total power saving is simulated for each parameter set.
There is a trade-off between the sampling ratio and power consumption that if $l$ is big, which means we compressed more sensors into small value of $m$, consumes more power than that using bigger $m$.

%%%%%%%%%%%%%%%%%%%%%%%%%%%%%%%%%%%%%%%%%%%%%%%%%%%%%%%%%%%%%%%%%%%%%%
% Conclusion

\vspace{-0.5em}
\section{Conclusion}
\label{sec:conclusion}
\vspace{-0.2em}
This paper presents a context-aware readout system for large TSPs 
equipped with an ultra-low-power always-on event or ROI detector.
Taking advantage of touch event sparsity, a novel sensing matrix and a simple ROI detection algorithm that can identify the activated sensor regions with small number of measurements are proposed.
Instead of turning on the readout circuits all the time, it allows the overall system
to save huge amount of energy by turning off the readout circuits when none of the sensors are activated, improving the energy efficiency by up to {42$\times$} over conventional always-on readout systems.

\vspace{-0.8em}
\section*{Acknowledgment}
The EDA Tool was supported by the IC Design Education Center (IDEC), Korea.

%%%%%%%%%%%%%%%%%%%%%%%%%%%%%%%%%%%%%%%%%%%%%%%%%%%%%%%%%%%%%%%%%%%%%%
% Reference
\vspace{-0.8em}
\bibliographystyle{IEEEtran}

\end{document}